\documentstyle[amsmath,amssymb,graphicx]{article}

\def\be{\begin{eqnarray}}
\def\ee{\end{eqnarray}}
\def\nn{\nonumber}

\def\tr{{\rm tr}\,}

\hoffset= - 1.1in         

\begin{document}

\hfill ITEP/TH-02/12

\bigskip

\bigskip

\bigskip

\centerline{\Large{Challenges of $\beta$-deformation
}}

\bigskip

\centerline{A.Morozov}

\bigskip

\centerline{\it ITEP, Moscow, Russia}

\bigskip

\centerline{ABSTRACT}

\bigskip

{\footnotesize
A brief review of problems, arising in the study of the
beta-deformation, also known as "refinement",
which appears as a central difficult element
in a number of related modern subjects:
$\beta \neq 1$ is responsible for deviation from free fermions
in $2d$ conformal theories, from symmetric omega-backgrounds
with $\epsilon_2=-\epsilon_1$
in instanton sums in $4d$ SYM theories, from eigenvalue matrix
models to beta-ensembles, from HOMFLY to super-polynomials
in Chern-Simons theory, from quantum groups to elliptic
and hyperbolic algebras etc.
The main attention is paid to the context of AGT relation
and its possible generalizations.
}

\bigskip

\bigskip

\section{ $\beta$-deformation}

This paper is a brief summary of fresh results, concerning the
new role of $\beta$-deformation -- an old subject
in the theory of matrix models and symmetric functions.
Today $\beta$-deformation is attracting increasing attention
because of its role in modern topics of theoretical physics,
including AGT relation \cite{AGT1},
AMM/EO topological recursion \cite{AMMsf,EO},
knot (Chern-Simons) theory \cite{CSt}-\cite{DMMSS} and matrix models \cite{betamamo}.
In what follows only direct references to recent statements
are given, all basic citations can be found in those papers.

\bigskip

The simplest example, explaining why $\beta$-deformation should naturally appear,
comes from elementary combinatorics.
Dedekind function counts Young diagrams with a given number of boxes:
if $\phantom._2\#_k$ is a number of integer partitions of the number $k$, then
\be
\sum_{k=0}  \phantom._2\#_k\, q^k = 1 + q + 2q^2 + 3q^3 + 5q^4 + 7q^5 + 11q^6 + \ldots
= \prod_{k=1}\ (1-q^k)^{-1}
\ee
Similarly McMahon formula counts the number of $3d$ Young diagrams:
\be
\sum_{k=0}\phantom._3\#_k\, q^k = 1 + q + 3q^2 +   \ldots
= \prod_{k=1}\ (1-q^k)^{-k}
\ee
However, this formula does not look like describing a
decomposition into elementary constituents, and it should be {\it refined} to
\be
\prod_{i,j}\ (1-q^it^j)^{-1}
\ee
Then McMahon formula arises at the special point $t=q$.
Thus conversion to elementary constituents often requires introduction of
an extra parameter in partition function -- transition from single to double
expansions, and in too many cases this happens in one-and-the-same
group-theory-related manner, allowing one to speak about
"double quantization".
This class of similar looking examples, coming from {\it a priori} different
subjects calls for development of a general theory of this peculiar
"$\beta$-deformation.
Parameter $\beta$ appears in $(q,t)$ parametrization through $t=q^\beta$,\
and undeformed situation corresponds to $\beta=1$.
The limits $\beta=0$ and $\beta=\infty$, as well as $q\rightarrow 0$
with $\beta$ fixed are non-trivial and always interesting.

In what follows we present various examples of how and where the $\beta$-deformation
appears in modern studies and try to emphasize the common features of these
examples and their intimate interrelations.

\subsection{Group theory}

The most effective is still the description of $\beta$-deformation
in terms of symmetric functions.
In this context one simply substitutes the Shur functions $S_R\{p_k\}$,
which are characters of the $GL(\infty)$ algebra
by MacDonald polynomials $M_R\{p_k\}$ \cite{McD}.

Shur functions are eigenfunctions of cut-and-join operators
$\hat W(\Delta)$,
\be
\hat W(\Delta) S_R = \varphi_R(\Delta)S_R
\ee
where $\Delta = \{\delta_1\geq\delta_2\geq\ldots\geq 0\}$ are integer partitions
or Young diagrams.
If infinitely many time variables $\{p_k\}$ are expressed through the
$N\times N$ matrix $X$ by {\it Miwa transformation}  $p_k = \tr X^k = kt_k$,
then Shur functions become $GL(N)$ characters $S_R[X] = s_R\{p_k=\tr X^k\}$, and
\be
\hat W(\Delta) = \ :\prod_i{\rm tr} \left(X\frac{\partial}{\partial X}\right)^{\delta_i}:
\ee
(normal ordering implies that all $X$-derivatives are standing to the right of all $X$'s.

Likewise, MacDonald polynomials are
eigenfunctions of the Ruijsenaars Hamiltonians \cite{RujsH}.

\subsection{Reductions}

Ruijsenaars integrable system is a generalization of Calogero one,
thus MacDonald polynomials are generalizations of the ordinary
Shur functions, which are the eigenfunctions of
Calogero Hamiltonians.
On the way there are two important intermediate cases:
Hall-Littlewood and Jack polynomials,
arising from MacDonald for $q=0$ and $q=t=0$ with $t=q^\beta$
respectively:

$$\begin{array}{ccccc}
&& {\rm algebra}  && \mbox{{\footnotesize Calogero}}\ \ \ \ \ \ \\
&& {\rm (Shur\ fns)} && \\
&   & & \nwarrow & {\footnotesize \phantom.^{\beta=1}}\ \ \ \ \ \ \ \ \ \ \ \ \ \ \ \ \\
{\rm quantum\ algebra} & &\ \ \ \uparrow\  {\footnotesize \phantom.^{t=q}} & &\beta-{\rm ensemble} \\
{\rm (Hall-Littlewood\ pols)} & & & &{\rm (Jack\ pols)}\\
\ \ \ \ \ \ \ \ \ \ \ \ \ \ \ \ \ \ \ \ \ \ \ \ \ \ \ \ {\footnotesize \phantom._{q=0}}& \nwarrow
& & \nearrow & {\footnotesize \phantom._{t,q\longrightarrow 1}}\ \ \ \ \ \ \ \ \ \ \ \ \ \\
&&{\rm MacDonald\ pols}&& \\
\ \ \ \ \ \ \ \mbox{{\footnotesize Ruijsenaars}} && q, \ \ t = q^\beta && \\ \\
&& \ \ \downarrow\ {\footnotesize \phantom.^{t=q}} && \\
&&{\rm elem.symmetric\ pols}\ m_R &&
\end{array}
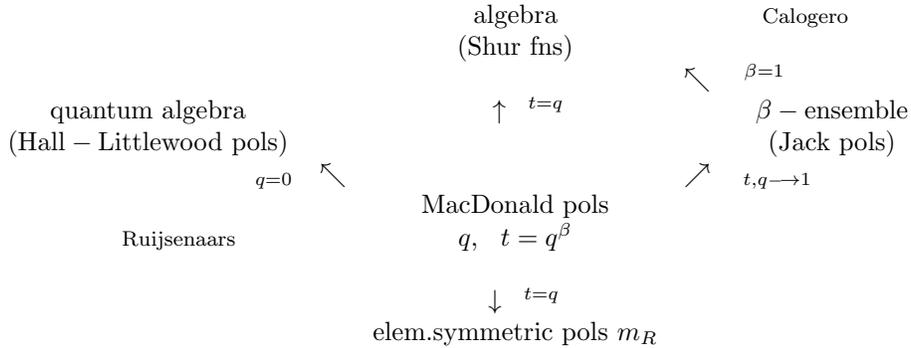
$$

\subsection{Orthogonal polynomials}

In matrix-model language this hierarchy of polynomials are associated with
the hierarchy of measures:
\be
{\rm Shur:} & \oint \prod_{i<j} (x_i-x_j)^2 \prod_i \frac{dx_i}{x_i} & \longrightarrow \nn\\
{\rm Jack:} & \oint \prod_{i<j} (x_i-x_j)^{2\beta} \prod_i \frac{dx_i}{x_i}
& \longrightarrow \nn \\
{\rm MacDonald:} & \oint \prod_{i \neq j} \prod_{k=0}^{\beta -1}(x_i-q^kx_j) \prod_i \frac{dx_i}{x_i}
\ee

Concrete role of these measures can be different, depending on the choice of
integration contours: for some choices the polynomials are orthogonal,
in more general case the linear and quadratic averages of polynomials
are fully-factorized Selberg-like integrals \cite{Selb}.

\subsection{Examples}

The simplest MacDonald polynomials are:

\be
M_1 = p_1 = {\rm Tr}\ X = \sum_i x_i, \ \ \ \ \ \ \ \ \ \ \ \ \ \
M_{11} = \frac{1}{2}\Big(-p\,_2 + p_1^2\,\Big) = \sum_{i< j} x_ix_j \nn \\ \nn \\
M_2 = \frac{1}{2}\left(\frac{(q-q^{-1})(t+t^{-1})}{qt-(qt)^{-1}}\,p\,_2
+ \frac{(q+q^{-1})(t-t^{-1})}{qt-(qt)^{-1}}\,p_1^2\right)\ \ \ \ \ \ \ \ \ \ \ \ \nn\\ \nn \\
\ldots \ \ \ \ \ \ \ \ \ \ \ \ \ \ \ \ \ \ \ \ \ \ \ \ \ \ \ \ \ \ \ \ \ \ \ \nn
\ee
Of special importance are MacDonald dimensions,
$M_R^* = M_R\{p=p^*\}$ the $\beta$-deformations of quantum dimensions:

$$p_k^* = \frac{A^k-A^{-k}}{t^k-t^{-k}} = \frac{\{A^k\}}{\{t^k\}}, \ \ \ \ \ \ A = t^N$$

$$M_1^* = \frac{A-1/A}{t-1/t} \ \ \ \stackrel{t=q}{\longrightarrow} \ \ [N]_q
\ \stackrel{q=1}{\longrightarrow}\  N$$
$$M_{11}^* = 
\frac{\{A/t\}\{A\}}{\{t\}\{t^2\}} \ \
 \ \stackrel{t=q}{\longrightarrow} \ \ \frac{[N-1]_q[N]_q}{[2]_q}
\ \stackrel{q=1}{\longrightarrow}\  \frac{(N-1)N}{2}$$
$$M_2^* = \frac{\{A\}\{Aq\}}{\{t\}\{qt\}}\ \
\ \stackrel{t=q}{\longrightarrow} \ \ \frac{[N]_q[N+1]_q}{[2]_q}
\ \stackrel{q=1}{\longrightarrow}\  \frac{N(N+1)}{2}$$
$$\ldots $$
They are given by the hook formula:
$$
M_R^* = \prod_{(i,j)\in R} \frac{\{Aq^{i-1}/t^{j-1}\}}{\{q^kt^{l+1}\}}
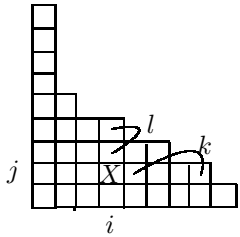
$$
\centerline{
\unitlength 1mm 
\linethickness{0.4pt}
\ifx\plotpoint\undefined\newsavebox{\plotpoint}\fi 
\begin{picture}(40,35)(30,25)
\put(31.849,31.008){\line(0,1){27.014}}
\put(31.849,58.022){\line(1,0){3.048}}
\put(34.897,58.022){\line(0,-1){27.014}}
\put(34.897,31.008){\line(-1,0){3.048}}
\put(31.849,54.868){\line(1,0){3.0482}}
\put(31.954,51.715){\line(1,0){2.838}}
\put(34.897,31.113){\line(1,0){24.071}}
\put(58.968,31.113){\line(0,1){3.048}}
\put(58.968,34.161){\line(-1,0){27.119}}
\put(37.42,30.903){\line(0,-1){.2102}}
\put(37.63,31.113){\line(0,1){15.031}}
\put(37.63,46.144){\line(-1,0){5.6761}}
\put(31.954,36.999){\line(1,0){23.545}}
\put(55.499,36.999){\line(0,-1){5.781}}
\put(31.954,39.837){\line(1,0){18.079}}
\put(50.033,31.218){\line(0,1){8.514}}
\put(52.661,36.684){\line(0,-1){5.466}}
\put(32.059,42.886){\line(1,0){11.983}}
\put(44.042,42.886){\line(0,-1){11.878}}
\put(46.985,39.417){\line(0,-1){8.304}}
\put(40.678,42.57){\line(0,-1){11.352}}
\put(31.849,48.877){\line(1,0){2.9431}}
\put(42.15,28.906){\makebox(0,0)[cc]{$i$}}
\put(29.326,35.738){\makebox(0,0)[cc]{$j$}}
\put(42.15,35.528){\makebox(0,0)[cc]{$X$}}
\qbezier(45.303,35.528)(55.972,41.677)(54.238,35.423)
\qbezier(42.465,38.261)(49.613,42.728)(42.465,41.519)
\put(54.763,39.417){\makebox(0,0)[cc]{$k$}}
\put(47.616,42.15){\makebox(0,0)[cc]{$l$}}
\end{picture}
}
where $\{z\} = z-z^{-1}$.
They should look familiar for those, who know Nekrasov functions \cite{Nef}
or topological vertex formulas \cite{TopV,Sulk}.
At $\beta\neq 1$  only $M_{11\ldots 1}^*$
are polynomials for $A=t^N$.

\subsection{The role of $\beta$-deformation}

Today $\beta$-deformation is finally in the mainstream:
it is no longer a game of mind,
it {\it appears} naturally in our theories.
Just two examples:

\bigskip

$\bullet$ {\bf AGT} \cite{AGT1,AGT2}

The starting point is a $6d$ conformal field theory compactified on a Riemann surface:
and it relates something $2d$, what lives on the Riemann surface,
with something in $4d$, which lives in uncompactified dimensions.
For example,

\bigskip

conformal blocks \cite{CFT} = LMNS integrals \cite{LMNS}
\be
c = (N-1)\left\{1 - N(N+1)\left(\sqrt{\beta} - \frac{1}{\sqrt{\beta}}\right)^2\right\}
\ee

$$ g_s = \sqrt{-\epsilon_1\epsilon_2} $$
$$\beta = -\epsilon_2/\epsilon_1  = b^2$$

\bigskip

LMNS integral = $\sum_{R_1,\ldots,R_N}$ Nekrasov functions.
Nekrasov functions have typical hook-product form,
similar to MacDonald dimensions

 \bigskip

$\bullet$ {\bf 3d AGT}

This is still a hypothetical relation, looked for in different directions \cite{3dAGT1}.
The most interesting version
involves $3d$ Chern-Simons theory, and relates, for example, the
S-duality (modular) transformations with
{\bf knot invariants}.

\bigskip

Wilson average in CS theory = HOMFLY polynomial
of two variables: $q = e^{2\pi i/(k+N)}$ and $a = q^N$

\bigskip

In Hikami representation through quantum dilogarithms \cite{Hik}
HOMFLY polynomials resemble Teshner's formulas for modular transforms \cite{Tesh},
but exact matching is somewhat problematic \cite{3dAGT2}

\bigskip

After $\beta$-deformation
HOMFLY$(a|q)$ $\ \stackrel{\beta \neq 1}{\longrightarrow}\ $ superpolynomial $P(A|q|t)$


\be
P_R[K](A|q|t) = \sum_{Q \vdash b[K]} c_R^Q[K] M_Q^*
\label{HYexpan}
\ee
where $K$ is a knot, $t = q^\beta$ and
only the $\beta$-quantum (MacDonald) dimensions $M_Q^*$ depend on $A = t^N$

\bigskip

In general expansion (\ref{HYexpan}) is ambiguous, ambiguity is fixed
by the quantum-$R$-matrix representation \cite{TR,MoSm}

\bigskip

After that (\ref{HYexpan}) can be used to extend (super)polynomials
to entire space of time-variables, by substituting $M_Q^* \longrightarrow M_Q\{p\}$
\cite{knMMM12}

\bigskip

Coefficients $c_R^Q[K]$ depend on the knot, they are rational functions of $q,t$,
and for toric knots are described by a simple $W$-representation.

\subsection{What survives after $\beta$-deformation?}
Everything, related to character calculus:

\bigskip

$\bullet$
Seiberg-Witten (SW) equations
\be
\left\{ \begin{array}{c}
a_i = \oint_{A_i} \Omega \\ \\
\frac{\partial \log Z}{\partial a_i} = \oint_{B_i} \Omega
\end{array} \right. \\ \nn
\ee
($\Longrightarrow\ $ quasiclassical integrability and WDVV equations)

$\bullet$ Virasoro constraints
$\ \longrightarrow \ $ AMM/EO topological recursion

$\bullet$ W-representations

$\bullet$ AGT relations

$\bullet$ knot invariants

\subsection{What is lost (modified) after $\beta$-deformation?}


Everything, related to KP-integrability:
\bigskip

$\bullet$ $Z = \tau$-function

$\bullet$ determinantal representations

$\bullet$ Harer-Zagier recursion

$\bullet$ Kontsevich matrix models

$\bullet$ Turaev-Reshetikhin construction

\bigskip

\noindent
In all cases $\beta$-deformed modifications are supposed to exist,
but are not yet available.

\subsection{Are there nice and natural decompositions?}

One of the puzzles with $\beta$-deformation is that
for $\beta\neq 1$
the natural quantities are no longer the ones with the most simple
algebraic properties:

$\bullet$ AGT could be a Hubbard-Stratanovich (HS) duality,
but Nekrasov functions are not the HS-duals of conformal blocks,
moreover, they have extra poles \cite{HSdua}.

$\bullet$ Naive link invariants for non-fundamental representations
$R\neq [1^{|R|}]$ are not the colored superpolynomials \cite{DMMSS}.

$$\begin{array}{|c||c|c|}
\hline
&{\rm natural\ quantities} & {\rm factorizable\ constituents} \\
\hline
{\rm Dotsenko-Fateev\ integral} &  {\rm Selberg\ correlators} & {\rm Nekrasov\ functions} \\
\hline
{\rm link\ invariants} & {\rm superpolynomials} & {\rm MacDonald\ dimensions} \\
\hline
\end{array}
$$

\bigskip

\newpage
\section{Matrix Models}

Matrix model $\tau$-functions play especially important role on string theory,
making them the most prominent candidates for the next generation of
special functions \cite{AMMsf}.

Eigenvalue matrix models possess a number of different definitions \cite{UFN3}:

\subsection{Multiple integrals (over eigenvalues)}

\be
Z = \left(\prod_{i=1}^N\int  e^{V(x_i)/g_s}dx_i\right) \Delta^2\{x\}
\ee
Exact evaluation of such integrals will one day be possible
in the context of {\bf non-linear algebra} \cite{NLA}.
The simplest available examples are  {\it integral\ discriminants} \cite{intdisc}:
$$\int\!\!\int dx dy\ e^{ax^2+bxy +dy^2} \sim \frac{1}{\sqrt{4ad-b^2}} = D_{2|2}^{-1/2}$$
$$\int\!\!\int  dx dy\ e^{ax^3+bx^2y+cxy^2 +dy^3} \sim {D_{2|3}^{-1/6}}$$
$$D_{2|3} = 27a^2d^2-b^2c^2-18abcd+4ac^3+4b^3d$$
In general ordinary discriminants only control
singularities of integral discriminants, but the answers are more involved.

While exact formulas of this type are not yet available in general,
there are many  other approaches,
which reveal a lot of hidden structures.

\subsection{Ward identities}

They are also known as  loop equations or Virasoro constraints \cite{loopeqs}-\cite{Virc}.
These are recursion relations for correlators
\be
\left(\sum_k kt_k\frac{\partial}{\partial t_{k+n}} + \sum_{a+b=n}
\frac{\partial^2}{\partial t_a\,\partial t_b}\right)Z = 0
\ee
They are preserved   by $\beta$-deformation,
only slightly modified \cite{betadefVC}.

\subsection{Integrable structure}

As a function of $t_k$ in $V(x) = \sum_k t_kx^k$,
the partition function
$Z\{t\}$ is a KP/Toda $\tau$-function \cite{GMMMO}
\be
\frac{\partial^2}{\partial t_1^2}\log Z_N = \frac{Z_{N+1}Z_{N-1}}{Z_N^2}
\ee
It is broken (at least essentially modified) by the $\beta$-deformation.

\subsection{Genus expansion}

$$\Delta\{x\} = \prod_{i\neq j} (x_i-x_j)^\beta$$

$$\log\Delta \ + \sum_i \frac{1}{g_s}V(x_i)
\ \sim\  N^2 \oplus N/g_s$$
\be
F = g_s^2\log Z = \sum_{p=0}^\infty g_s^{2p}F_p(a)
\ee
where $a$ is the t'Hooft coupling constant $a=Ng_s$.
If there are many different integration contours
over different eigenvalues, then there are many
parameters $a_I$ instead of a single $a$.

In perturbation theory $F_0$ is a sum of planar diagrams
and so on.

$\bullet$
Spectral curve $\Sigma$ is defined at the genus-zero level,
i.e. is hidden in $F_0$.
It plays a prominent role in two places:
resolvents and SW equations.

{\bf Resolvents} are peculiar generating functions of correlators
\be
\rho^{(p|m)}(z_1,\ldots,z_m) = \left<\prod_{i=1}^m {\rm Tr} \frac{dz_i}{z_i- X}\right>_p
= \sum_{\{k_i\}} \frac{1}{z_i^{k+1}} \left<\prod_i {\rm Tr} X^{k_i}\right>_p
\ee

\bigskip

Advantages of this definition are: \\

$\bullet$ Resolvents are meromorphic poly-differentials on $\Sigma$ \\

$\bullet$ As a consequence of {\bf Virasoro constraints},
they can be recursively reconstructed for a given $\Sigma$ with a
Seiberg-Witten differential
$\Omega^{(0)} = \rho^{(0|1)} \sim y(z)dz$ and Bergmann kernel $\rho^{(2|0)}$
by the AMM/EO recursion \cite{AMMsf,EO}.

\bigskip

However, it also has a drawback: \\

$\bullet$ sum over genera diverges, in particular

\be
\Omega(z) = \rho^{(\cdot|1)}(z) = \sum_p g_s^{2p}\rho^{(p|1)}(z)
\ee
can {\it not} be restored from the AMM/EO recursion.

\bigskip

At the same time $\Omega(z)$ is really important:
it is the SW differential for the free energy $F(a) = \sum_p g_s^{2p}F_p(a)$,
i.e. it enters the SW equations.

\subsection{SW equations}

This is a consistent system of equations
\be
\left\{ \begin{array}{c}
a_i = \oint_{A_i} \Omega \\ \\
\frac{\partial \log Z}{\partial a_i} = \oint_{B_i} \Omega
\end{array} \right.
\ee
which seems to be true both in matrix models and $\beta$-ensembles
(this is generally believed, but not proved).

The simplest is the Gaussian example with $\beta=1$ and the spectral curve
$\ \ \Sigma: \ \ y(z)^2 = z^2 - 4g_sN$ \cite{betadefoVC}
$$
Z_N = \frac{1}{N!}\int (x_i-x_j)^2 e^{-x_i^2/2g_s} dx_i \sim g_s^{N^2/2}\prod_{k=1}^{N-1} k!
$$
$$\frac{\partial}{\partial N}\sum_{k=0}^{N-1}f(k) = \sum_k \frac{B_k}{k!}\partial^kf(N)$$
$$
\frac{\partial}{\partial N}\log Z_N = N(\log g_sN -1) + \sum_k \frac{B_{2k}}{k}\frac{1}{N^{2k-1}}
$$

\bigskip

$$
\Omega(z) = -\frac{y(z)}{2} + \frac{g_s^2}{y(z)^5} + \frac{21g_s^4(z^2+g_sN)}{y(z)^{11}} + \ldots
$$
$$\oint_A \Omega(z) = N, \ \ \ \ \ \ \oint_B\Omega(z) = \frac{\partial}{\partial N}\log Z_N $$
In Gaussian case with $\beta =1$ these SW equations can be proved
from integrability  {\footnotesize [1011.5629]}.
However, SW equations remain true for $\beta\neq 1$, while integrability is broken
or at least modified.

Generalization to non-Gaussian case is provided by the theory of Dijkgraaf-Vafa phases
\cite{DV} in matrix models.

\subsection{How to define $\rho^{(\cdot|1)}$?}

A possible key here is the {\bf Harer-Zagier recursion}.
In variance with AMM/EO recursion \cite{AMMsf,EO}, implied by the Virasoro constraints \cite{Virc},
this one rather follows from integrability \cite{HZrec} -- and thus behave much worse under
the $\beta$-deformation.

\bigskip

Gaussian model ($V(x) = x^2/2$):

$$\rho(z) = \sum_k \frac{1}{z^{2k+1}}\left< {\rm Tr} X^{2k}\right>$$
$$\phi(t) = \sum_k \frac{t^{2k}}{(2k-1)!!}\left< {\rm Tr} X^{2k}\right>$$
$$e(s) = \sum_k \frac{s^{2k}}{(2k)!}\left< {\rm Tr} X^{2k}\right>$$

{\footnotesize
$$\left< {\rm Tr} X^{2k}\right>^{N=1} \sim (2k-1)!! \ \ \longrightarrow \ \
\left< {\rm Tr} X^{2k}\right>_0 \sim \frac{(2k-1)!!}{(k+1)!} \ \ ({\rm Catalan\ numbers})$$
}

\vspace{-0.3cm}

$$\phi(t|N) = \frac{1}{2t^2}\left(\left(\frac{1+t^2}{1-t^2}\right)^N-1\right)$$

\bigskip

$\bullet$ $N \longrightarrow \lambda$:
$$\hat \phi(t|\lambda) = \sum_{N=0}^\infty \phi(t|N)\lambda^N
= \frac{\lambda}{\lambda-1}\cdot\frac{1}{1-\lambda - (1+\lambda)t^2}$$

$\bullet$ multi-point correlators:
$$\hat\phi_{odd}(t_1,t_2|\lambda) = \frac{\lambda}{(1-\lambda)^{3/2}}
\frac{\arctan\frac{t_1t_2\sqrt{1-\lambda}}{\sqrt{1-\lambda + (1+\lambda)(t_1^2+t_2^2)}}}
{\sqrt{1-\lambda + (1+\lambda)(t_1^2+t_2^2)}}$$

$\bullet$ other generating functions:

$$\hat e(s|\lambda) = \frac{\lambda}{(1-\lambda)^2} e^{\frac{1+\lambda}{1-\lambda}s^2}$$
$$\hat\rho(z|\lambda) = \frac{i\lambda}{(1-\lambda)\sqrt{1-\lambda^2}}\
{\rm erf}\left(iz\sqrt{\frac{1-\lambda}{1+\lambda}}\right) =$$
$$ =\sum_{k=0}^\infty \frac{\lambda(1+\lambda)^k}{(1-\lambda)^{k+2}}\frac{(2k-1)!!}{z^{2k+1}}$$
$$\Longrightarrow \ \ \rho(z) = \frac{z-y(z)}{2} + \frac{N}{y^5(z)} + \frac{21N(z^2+N)}{y^{11}(z)}
+ \ldots$$

\bigskip

$\bullet$ $\beta$-deformation:\\
for $\beta=2,1/2$ -- 1-point functions are still expressed through $\arctan$\\
for $\beta=3$ -- a differential equation can be written for the 1-point function.

\subsection{W-representations \cite{Wreps}}

Partition functions can be considered as
a result of "evolution", driven by cut-and-join (W) operators
from very simple "initial conditions"

\be
Z\{p\} = e^{g\hat W} \tau_0\{p\}
\ee

\bigskip

If $W \in UGL(\infty)$, then KP/Toda-integrability is preserved

\bigskip

\be
\hat W_n = \frac{1}{2}\sum_{a,b} \left((a+b+n)p_ap_b\frac{\partial}{\partial p_{a+b+n}}
+ abp_{a+b-n}\frac{\partial^2}{\partial p_a\partial p_b}\right)
\ee

$\bullet$ Hermitian matrix model $\ \ Z_N = \int dX e^{\sum_k \frac{p_k}{k}{\rm Tr} X^k}$
\be
Z_N = e^{\hat W_{-2}} e^{Np_0}
\ee

$\bullet$ Kontsevich model \cite{alWrep}
$\ \ Z = \int dX e^{{\rm Tr}(\frac{1}{3} X^3 - L^2X)}$, $p_k = {\rm Tr}L^{-k}$

\be
Z = e^{\hat W_{-1}^K}\cdot 1
\ee \ \ \ \ \ \ \ \ \  \ \ \ \ \ \
\vspace{-0.3cm}
$${\rm where} \ \ \ \ \ \hat W_{-1}^K = \frac{2}{3}\sum \left(k+\frac{1}{2}\right)\tau_k L^K_{k-1}$$

$\bullet$ Hurwitz model \cite{BoMa}
\be
Z = e^{t\hat W_0} e^{p_1}
\ee

$\bullet$ Toric knots and links \cite{torWrep,DMMSS}
\be
Z = q^{\frac{n}{m}\hat W_0} \prod_{{\rm link\ comps}} \tilde\chi_R
\ee

These formulas might imply a striking relation between
Hurwitz and torus-knot theories.

\newpage
\section{AGT relations \cite{AGT1,AGT2}}

The main subjects in this story are:

$\bullet$ Dotsenko-Fateev matrix model \cite{DFmamo}

$\bullet$ Hubbard-Stratanovich duality \cite{HSdua}

$\bullet$ Relation to integrable systems \cite{GKMMM,NS1}

$\bullet$ Bohr-Sommerfeld integrals \cite{BSint}

\bigskip

\noindent
The main fact is that
certain non-trivial quantities in four different classes of theories
are currently known to be the same:

$$
\begin{array}{ccc}
{\cal N}=2\ {\rm SYM\ models} & \stackrel{{\rm AGT}}{\longleftrightarrow} &
2d\ {\rm CFT\ conformal blocks} \\ &&\\
\updownarrow\ & \!\!\!\!\!\!\!\!\!\!\!\!\!\!\!\!\!\!\!\!\!\!\!\!
{\rm dictionary\ [1995-97]} & \updownarrow \\ &&\\
1d\ {\rm integrable\ systems} & \stackrel{?}{\longleftrightarrow}& {\rm DF/Penner\ matrix\ model}\\
&&
\end{array}
$$

\noindent
Each of the entries can be used to label universality classes.
Each type of theories implies certain "natural" deformations
and generalizations, and the question is what it corresponds to
in the other corners of the table.

\subsection{Quantization of integrable systems}

The left vertical arrow is well known in
the original Seiberg-Witten context \cite{SW}:
the SW equations for the non-perturbative (instanton-induced) prepotential
in $N=2$ SYM theories
are expressed through the action variables for classical
$1d$ integrable systems \cite{GKMMM}.

When SYM theory is deformed by the $\Omega$-background \cite{LMNS},
with two parameters $\epsilon_1$ and $\epsilon_2$,
integrable system is deformed.

The first natural guess is that the deformation is just
the quantization of integrable system.
The Shroedinger-like equations arise as Fourier transforms of the Baxter equations,
and they are AGT-related to the equations for conformal correlators with
insertions of degenerate states.

SW description appears through the  Bohr-Sommerfeld integrals \cite{BSint}:
if the wave function
$\Psi(z) = \exp\int^z \Omega$, where $\Omega = Pdz$
with a quantum-corrected momentum $P(z)$, then
$\frac{\partial F}{\partial a} = \oint_B\Omega, \ \ \ a = \oint_A\Omega$.

But in fact this quantization of integrable system
is associated with Nekrasov-Shatashvili (NS) limit
$\epsilon_1\rightarrow 0$,\ $\beta\rightarrow \infty$.

\subsection{Matrix-model representation of conformal blocks \cite{DFmamo}}

Conformal block is parameterized by the following data
(the 4-point example):

\centerline{
\begin{picture}(100,60)(-20,-30)
\put(0,0){\line(1,0){60}}
\put(0,0){\line(-2,1){30}}
\put(0,0){\line(-2,-1){30}}
\put(60,0){\line(2,1){30}}
\put(60,0){\line(2,-1){30}}
\put(-50,-10){\makebox(0,0)[cc]{$V_{\alpha_1}(0)$}}
\put(-50,18){\makebox(0,0)[cc]{$V_{\alpha_2}(q)$}}
\put(110,18){\makebox(0,0)[cc]{$V_{\alpha_3}(1)$}}
\put(110,-10){\makebox(0,0)[cc]{$V_{\alpha_4}(\infty)$}}
\put(30,5){\makebox(0,0)[cc]{$\alpha$}}
\end{picture}
}

\noindent
It is actually equal to
\be
\left< e^{\alpha_1\phi(0)} e^{\alpha_2\phi(q)} e^{\alpha_3\phi(1)} e^{\alpha_4\phi(\infty)}
\ \prod_{i=1}^{N_1}\int_0^q e^{b\phi(x_i)} \ \prod_{j=1}^{N_2} \int_0^1 e^{b\phi(y_j)}
\right>\ =
\ee
$$\alpha_1+\alpha_2 + bN_1 = \alpha$$
$$\alpha + \alpha_3+\alpha_4 + bN_2 = 0$$

$$
\!\!\! =\!\int\!\! dx_i \!\int\!\! dy_j\, (x_i-x_{i'})^{2\beta}(y_j-y_{j'})^{2\beta}
\underline{(x_i-y_j)^{2\beta}}
(x_iy_j)^{2\alpha_1b}\big((q-x_i)(q-y_j)\big)^{2\alpha_2b}\big((1-x_i)(1-y_j)\big)^{2\alpha_3b} =
$$
\be
= \int_{d\mu (x)} \int_{d\mu(y)}\ \Big({\rm Mixing \ term}(x|y)\Big)^2
\label{Mixterm}
\ee
For $\beta=1$
\be
d\mu(x) = \prod_{i<i'} (x_i-x_{i'})^{2} \prod_i x_i^a (1-x_i)^c dx_i
\ee
is Selberg measure.
Natural are Selberg averages of Shur functions,
they are nicely factorized -- and they are exactly the Nekrasov functions.

\bigskip

$\beta$  deformation implies that the measure is changed for
MacDonald one:
\be
\int_{Jackson} \prod_{k=0}^{\beta-1} \prod_{i\neq i'}(x_i - q^kx_{i'})
\ee
$$q^\beta = t$$
However, with this measure the averages of Jack and MacDonald polynomials
are often {\it not} factorized, instead they
linearly decompose into factorizable quantities (Nekrasov functions)

\subsection {Pure gauge limit and BGW model}

This limit -- natural from the perspective of the SYM models --
corresponds to breakdown of conformal invariance.
In this "pure gauge" limit the logarithmic $\beta$-ensemble
turns into the BGW model
(for $\beta=1$ in involves unitary rather than Hermitian matrices!)
\cite{BGWmod}.

Elliptic case, associated with the toric conformal blocks
is supposedly related to double-cut BGW \cite{DVdc}.

BGW model is an important building block
in M-theory of matrix models \cite{BGWmatrix}.

\subsection{Conformal block as average of characters}

Continuing from (\ref{Mixterm}), conformal block is equal to \cite{IO}
$$
\approx \int_{d\mu (x)} \int_{d\mu(y)}\ \exp\left(2\beta\sum_{i,j}\log(1-x_iy_j)\right) =
$$
$$
=\int_{d\mu (x)} \int_{d\mu(y)}\ \exp\Big(2\beta \sum_k \frac{p_k\bar p_k}{k}\Big) =
$$
$$
= \int_{d\mu (x)} \int_{d\mu(y)}\ \Big(\sum_A\chi_A(X)\chi_A(Y)\Big) \Big(\sum_B\chi_B(X)\chi_B(Y)\Big) =
$$
\be
= \sum_{A,B}\left( \int_{d\mu (x)}\chi_A(X)\chi_B(X)\right) \left(\int_{d\mu(y)}\chi_A(Y)\chi_B(Y)\right)
\ee
Here
$$ p_k = {\rm Tr} X^k, \ \ \ \ \bar p_k = {\rm Tr} Y^k$$
and for arbitrary $\beta$ the role of characters is played by MacDonald polynomials,
arising from
\be
\exp \sum_k \frac{[\beta]_{q^k} p_k\bar p_k}{k} = \sum_A \frac{C_A}{C_{A'}} M_A(X)M_A(Y)
\ee

\subsection{AGT as Hubbard-Stratanovich duality \cite{HSdua}}

\vspace{0.5cm}

\centerline{
\begin{picture}(100,50)(40,0)
\put(0,0){\line(1,0){40}}
\put(0,0){\line(-2,1){20}}
\put(0,0){\line(-2,-1){20}}
\put(40,0){\line(2,1){20}}
\put(40,0){\line(2,-1){20}}
\put(170,-10){\line(0,1){20}}
\put(170,10){\line(-2,1){20}}
\put(170,-10){\line(-2,-1){20}}
\put(170,10){\line(2,1){20}}
\put(170,-10){\line(2,-1){20}}
\put(110,0){\makebox(0,0)[cc]{=}}
\put(-30,25){\makebox(0,0)[cc]{$\chi_A(X)$}}
\put(-30,-25){\makebox(0,0)[cc]{$\chi_A(Y)$}}
\put(70,25){\makebox(0,0)[cc]{$\chi_B(X)$}}
\put(70,-25){\makebox(0,0)[cc]{$\chi_B(Y)$}}
\put(140,35){\makebox(0,0)[cc]{$\chi_A(X)$}}
\put(140,-35){\makebox(0,0)[cc]{$\chi_A(Y)$}}
\put(200,35){\makebox(0,0)[cc]{$\chi_B(X)$}}
\put(200,-35){\makebox(0,0)[cc]{$\chi_B(Y)$}}
\end{picture}
}

\bigskip

\bigskip

\bigskip

\bigskip

\bigskip

\bigskip

\be{\footnotesize
\sum_{X,Y} \left(\sum_A \chi_A(X)\chi_A(Y)\right) \left(\sum_B \chi_B(X)\chi_B(Y)\right) =
  \sum_{A,B} \left(\sum_X \chi_A(X)\chi_B(X)\right) \left(\sum_Y \chi_A(Y)\chi_B(Y)\right)
}\ee


\be
{\rm Conformal\ block}\ =\ \sum_{A,B} N_{A,B}
\ee

\be
\int_{d\mu(X)} \chi_A(X)\chi_B(X) \int_{d\mu(Y)} \chi_A(Y)\chi_B(Y)
 \ \stackrel{?}{=}\ N_{A,B}
\ee
This relation is true for $\beta=1$.

\subsection{The extra poles problem for $\beta\neq 1$  \cite{HSdua}}

However, things are {\it not} so simple for $\beta\neq 1$.
Already in the simplest example,
$$
<\chi_{[1]}\ \chi_\bullet><\chi_{[1]}\ \chi_\bullet>
+ <\chi_\bullet\ \chi_{[1]}><\chi_\bullet\ \chi_{[1]}>\ =
$$
$$
=\frac{1}{(z-\epsilon)}\frac{1}{(z+\epsilon)} + \frac{1}{(z+\epsilon)}\frac{1}{(z-\epsilon)}=
$$
$$
= \frac{2 }{z^2-\epsilon^2}
= \frac{1}{z(z-\epsilon)} + \frac{1}{z(z+\epsilon)} = N_{[1],\bullet} + N_{\bullet,[1]}
$$
particular Nekrasov functions
for $\epsilon\neq 0$ ($\beta\neq 1$)
have extra zeroes (at $z = 0$), not present in Kac determinant,
i.e. not present in conformal blocks.

Instead Nekrasov functions are nicely factorized,
while Selberg correlators for $\beta\neq 1$ are not:

$$<\chi_{[3]}\ \chi_\bullet>_{BGW}\ \sim\
z^2-(5\epsilon_1+8\epsilon_2)z+6\epsilon_1^2+23\epsilon_1\epsilon_2+19\epsilon_2^2$$
$$\stackrel{\epsilon_2=-\epsilon_1}{\longrightarrow}\
z^2 + 3\epsilon_1 z + 2\epsilon_1^2 = (z+\epsilon_1)(z+2\epsilon_1)$$

The natural quantities, e.g. the Selberg correlators
(involved into duality relations)
are non-trivial linear combinations of
the nicely factorized functions (Nekrasov functions),
which possess extra singularities

Similar is the situation with knot invariants:
superpolynomials for unknots (the natural quantities)
are linear combinations of
MacDonald dimensions (the nicely factorized quantities).

\newpage

\section{Knots}

The story is not so much about knots --
rather again about the averages of characters.
Usually we are interested in the following chain
of objects and quantities \cite{knMMM12}:

\bigskip

knot $\ \longrightarrow \ $ Wilson average
$\ K=\left< {\rm Pexp} \oint_{{\rm knot}} {\cal A}\right>_{CS}$
$\ \longrightarrow \
K\{\bar p \,|\, {\rm knot} \} =  \sum_R K_R({\rm knot})\chi_R\{\bar p\}
\ \longleftrightarrow\  \tau\{\bar p\,|G\}$

\bigskip
\noindent
Wilson average $K$ is a polynomial of $q$ and $A=q^N$ (called
HOMFLY polynomial), and
$\tau$ at the end should be some kind of
generalized $\tau$-function
and $G$ -- a point of the universal moduli space
(universal Grassmannian).
Situation should be similar to matrix models, where
different models are labeled by different $G$.
Likewise here different knots could also be labeled by different $G$.
However, for this to work some still unknown modification of
KP/Toda $\tau$-function is needed.

\subsection{"Special" polynomials \cite{DMMSS}}

At $q=1$ HOMFLY polynomials are reduced to
\be
S_R(A) = \Big(S_{[1]}(A)\Big)^{|R|}D_R
\ee
Coefficients of these "special"  polynomials $S(A)$ are Catalan-like numbers,
counting the numbers of certain paths on $2d$ lattices.
They satisfy Pl\"ucker relations and thus provide KP $\tau$-functions
\be
\tau\{p\} = \sum_R S_R(A)\chi_R\{p\}
\ee
However, for $q\neq 0$ this is no longer true
and $\tau$-function should be deformed away from KP (free-fermion) locus.

\subsection{$\beta$-deformation of HOMFLY polynomials}

For a given knot $K$ and representation (Young diagram) $R$

$$
\begin{array}{cccc}
&& \!\!\!\!\!\!\!\!\!\!\!\!\!\!\!\!\!\!\!\!\!\!\!\!\!\!\!\!\!\!\!\!\!\!\!\!
{\rm Superpolynomial}\ P_R(A|q|t) &\\
&&&\\
&\swarrow t\approx q &&
\!\!\!\!\!\!\!\!\!\!\!\!\!\!\!\!\!\!\!\!\!\!\!\!\!\!\!\!\!\!\!\!\!\!\!\!\searrow A\approx 1 \\
&&&\\
\!\!\!\!\!\!\!\!\!\!\!\!\!\!\!\!\!\!CS\ \longrightarrow &
 \!\!\!\!\!\!\!\!\!\!\!\!\!\!\!\!\!\!\!\!\!\!\!\!\!\!\!
\boxed{{\rm HOMFLY}\ H_R(A|q)} &&
\!\!\!\!\!\!\!\!\!\!\!\!\!\!\!\!\!\!\!\!\!\!\!\!\!\!\!\!\!\!\!\!\!\!\!\!
 {\rm Heegard-Floer}\ HF_R(q|t)  \\
&&&\\
q=1\swarrow & \ \ \ \ \ \ N=2\searrow N=0 &&
\!\!\!\!\!\!\!\!\!\!\!\!\!\!\!\!\!\!\!\!\!\!\!\!\!\!\!\!\!\!\!\!\!\!\!\!\swarrow t\approx q\\
&&&\\
{\rm Special}\ S_R(A) & \!\!\!\!\!\!\!\!\!\!\!\!\!\!\!\!\!\!{\rm Jones}\ J_R(q) &
 \!\!\!\!\!\!\!\!\!\!\!\!\!\!\!\!\!\!\!\!\!
 {\rm Alexander}\ {\cal A}_R(q) &
\end{array}
$$

\bigskip

Here
$$A = t^N = q^{\beta N}$$
$q = \exp\frac{2\pi i}{k+N}$,\ \ \
and $A\sim \exp({\rm t'Hooft\ coupling})$ remains finite in the loop expansion.

\subsection{Some objects, presumably associated with knots}

Ideally one can look for at least the following mappings:

$$
{\rm knot} \ \longrightarrow\
\left\{ \begin{array}{c}
{\rm point\ of\ the\ universal\ Grassmannian\ (a\ dream ?)} \\ \\
{\rm vector\ in\ the\ Hilbert\ space},\\
{\rm  where\ modular\ operators}\ S\ {\rm and}\ T\ {\rm are\ acting}\\ \\
{\rm vector\ in\ the\ space\ of\ characters}\\
{\rm (quantum\ or\ Macdonald\ dimensions)}
\end{array}\right.
$$ \\
However, all of them still need to be accurately defined.

\subsection{$R$-matrix representation of HOMFLY polynomials  \cite{TR,MoSm}}

In the gauge $A_0=0$
knot invariants are described in terms of knot diagrams, so that:

$\bullet$
knot is substituted by  as a braid

$\bullet$
Element of a braid group is a product of quantum $R$-matrices
(somehow generalized after the $\beta$-deformation)

$\bullet$
K = "trace" of an element a braid group

\subsection{Torus knots $T[m.n]$}

Torus knots and links are made from a special braid element ${\cal R}_m$:
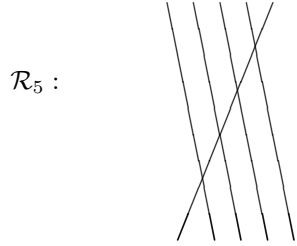
$$
\begin{picture}(100,80)(-40,-70)
\put(0,0){\line(1,-5){18}}
\put(10,0){\line(1,-5){18}}
\put(20,0){\line(1,-5){18}}
\put(30,0){\line(1,-5){18}}
\put(40,0){\line(-2,-5){36}}
\put(-50,-30){\makebox(0,0)[cc]{${\cal R}_5:$}}
\end{picture}
$$


\bigskip

$$H^{[m,n\,]}_R = {\rm Tr} \left({\cal R}_m\right)^n$$
where trace is defined so that
\be
{\rm Tr}_Q\ I^{\otimes m} = {\rm tr}_Q\ q^{\rho}
= \sum_{\vec\alpha\in Q} q^{\vec\rho\vec\alpha} = \chi_Q^*
= {\rm quantum\ dimension\ of\ representation\ } Q
\ee

Original representation is decomposed into irreducible ones:
\be
R_1\otimes \ldots \otimes R_m \ = \ {\oplus}_{Q\ \vdash\ (|R_1|+\ldots+|R_m|)}\ c_R^Q\,\cdot Q
\ee
The crucial fact is that
$Q$  are eigenspaces of the product ${\cal R}_m$
with certain eigenvalues $\lambda_Q$.
Therefore \cite{torWrep}
\be
H^{[m,n\,]}_R(A|q) \ \ =\ \  \sum_Q c^Q_R \lambda_Q^{\frac{n}{m}} \chi_Q^*
\ \ = \ \ e^{\frac{n}{m}\hat W_0} \left.\sum_Q c^Q_R \chi_Q\{p\}\right|_{p = p^*}
\ee

\subsection{MacDonald dimensions}

MacDonald dimensions ($\beta$-deformations of the usual quantum ones)
are defined as $M_R^* = M_R\{p=p^*\}$ with

$$p_k^* = \frac{A^k-A^{-k}}{t^k-t^{-k}} = \frac{\{A^k\}}{\{t^k\}}, \ \ \ \ \ \ A = t^N
\ \ \ \ \ \{z\} = z-1/z$$

$$M_1^* = \frac{A-1/A}{t-t/t} \ \ \ \stackrel{t=q}{\longrightarrow} \ \ [N]_q
\ \stackrel{q=1}{\longrightarrow}\  N$$
$$M_{11}^* = 
\frac{\{A/t\}\{A\}}{\{t\}\{t^2\}} \ \
 \ \stackrel{t=q}{\longrightarrow} \ \ \frac{[N-1]_q[N]_q}{[2]_q}
\ \stackrel{q=1}{\longrightarrow}\  \frac{(N-1)N}{2}$$
$$M_2^* = \frac{\{A\}\{Aq\}}{\{t\}\{qt\}}\ \
\ \stackrel{t=q}{\longrightarrow} \ \ \frac{[N]_q[N+1]_q}{[2]_q}
\ \stackrel{q=1}{\longrightarrow}\  \frac{N(N+1)}{2}$$
$$\ldots $$

\subsection{$W$-representation of HOMFLY polynomials for torus knots}


The relevant $W$-operator appears to be the simplest non-trivial
cut-and-join operator
\be
\hat W_0 =  \hat W[2] =
\sum_{a,b\geq 1} \left( (a+b)p_ap_b\frac{\partial}{\partial p_{a+b}}
+ abp_{a+b}\frac{\partial^2}{\partial p_a\partial p_b}\right)
\ee
for which Shur functions
$$s_1\{p\} = p_1,\ \ \ \
s_2\{p\} = \frac{1}{2}(p_2 + p_1^2), \ \ \ \
s_{11}\{p\} = \frac{1}{2}(-p_2 + p_1^2), \ \ \ldots
$$
are the common eigenfunctions:
\be
\hat W[2] s_Q\{p\} = \varkappa_Q s_Q\{p\},
\ \ \ \ \ \ \ \lambda_Q = q^{\varkappa_Q}
\ee
Eigenvalues of $\hat W[2]$ are especially simple
\be
\varkappa_Q = \sum_i q_i(q_i-2i+1) = \nu_Q - \nu_{Q^T} \nn\\
\nu_Q = \sum_i (i-1)q_i
\ee
For general theory of cut-and-join operators see \cite{MMN}.
In general eigenvalues depend on a pair of Young diagrams
and are essentially the generic characters of symmetric group $S(\infty)$.

\subsection{Initial conditions for the $n$-evolution}

These "initial conditions" 
are very simple, e.g.

\be
H_1^{[m,n]} = \left.q^{\frac{n}{m}\hat W[2]}\ p_m\right|_{p=p^*} \nn \\
H_R^{[m,n]} = \left.q^{\frac{n}{m}\hat W[2]}\ s_R\{p_{mk}\}\right|_{p=p^*}
\ee
for mutually prime $n$ and $m$, and
\be
H_{R_1\ldots R_m}^{[m,mk]} =  \left.q^{k\hat W[2]}\
s_{R_1}\{p_{mk}\}\ldots s_{R_m}\{p_{mk}\}\right|_{p=p^*}
\ee

In the latter case (fully disconnected link) they simply follow from
the fact that
$T[m,n]$ for $n=0$ is a set of $m$ unknots.

In the former case (connected knot) for $n=1$ there is a single unknot,
i.e. $H_R^{[m,1]} \sim s_R^*$.

\be
H^{[m,n\,]}_R(A|q)
\ \ = \ \ e^{\frac{n}{m}\hat W} \left.\sum_Q c^Q_R \chi_Q\{p\}\right|_{p = p^*}
 \ \ =\ \  \sum_Q c^Q_R q^{\frac{n}{m}\varkappa_Q} \chi_Q^*
\ee

\subsection{
Reformulation in terms of Frobenius algebra}

Frobenius algebra is a set of data:
\{linear space + multiplication + linear form\}, and
\be
H^{[m,n\,]}_R(A=q^N|q)
\ \ = \ \
\Big< s_R[U^m] \Big> = \sum_{Q} c_R^Q \Big< s_Q[U] \Big>
\ee

\be
\Big< s_Q[U] \Big> \sim q^{\frac{n}{m}\varkappa_Q}s_Q^*
\ee
Matrix-model realization of this linear form ($q=e^\hbar$) is \cite{BEM}
\be
\Big<F[U]\Big> = \int du_i e^{u_i^2/\hbar} \sinh  \sqrt{\frac{n}{m}}\frac{u_i-u_j}{2}
\sinh  \sqrt{\frac{m}{n}}\frac{u_i-u_j}{2} \
F\left[\exp\left(\sqrt{\frac{n}{m}}u_i\right)\right]
\ee

\subsection{Split W-representation for toric superpolynomials \cite{DMMSS}}

It is basically provided by straightforward
deformation from Shur to MacDonald functions:\\
\be
H_R^{[m,n\,]}(A|q) = \sum_Q c^Q_R q^{-\frac{n}{m}(\nu_Q-\nu_{Q'})} s_Q^*
\ \ \ \longrightarrow \ \ \
P_R^{[m,n\,]}(A|q|t) = \sum_Q c^Q_R \ q^{-\frac{n}{m}\nu_Q}\ t^{\frac{n}{m}\nu_{Q'}}\ M_Q^*
\label{torsup}
\ee
and have a form of a split (refined) W-representation

\subsection{How to choose the coefficients $c_R^Q$ \cite{DMMSS}?}

\noindent

$\bullet$ They depend on the series $T[m,mk+p]$,\ $p=0,1,\ldots,m-1$.

$\bullet$ They satisfy "initial conditions" at $k=0$:
$T[m,p] = T[p,m]$, $p<m$.

$\bullet$ They are such, that  $P^{[m,mk+p]}_R(A|q|t)$ is a polynomial
in all its variables with positive coefficients for all $k$ at once.

\bigskip

Initial condition would be sufficient, if imposed for all values
of time-variables $p_k$.
Actually it is imposed only on the subspace $p_k=p^*_k = \frac{A^k - A^{-k}}{t^k - t^{-k}}$,
and this is not sufficient for $|Q|\geq 4$.
The third condition should be used.
It is tedious, but it works:

$$p_m = \sum_{Q \vdash m} \bar c_{[1]}^Q M_Q\{p\}$$
$$ c_{[1]}^Q = \bar c_{[1]}^Q\cdot \gamma_{[1]}^Q$$

\bigskip

$$\gamma^{[2]} = \frac{1+q^2}{1+q^2} = 1, \ \ \ \ \gamma^{[11]} = \frac{1+t^2}{1+q^2}$$
$$\!\!\gamma^{[3]} = \frac{1+q^2+q^2q^2}{1+q^2+q^2q^2} = 1, \ \ \
\gamma^{[21]} = \frac{1+q^2 + q^2t^2}{1+q^2 + q^2q^2}, \ \ \
\gamma^{[111]} = \frac{1+t^2 + t^2t^2}{1+q^2 + q^2q^2}$$

\bigskip

\noindent
General formulas can be also worked out for other series,
but they look better after additional structures are revealed,
see the second paper in ref.\cite{DMMSS}.

\subsection{Evidence in favor of the answer (\ref{torsup}) for torus superpolynomials }

\noindent

$\bullet$ It is consistent with all known superpolynomials
in all fundamental representations $R = [1^{|R|}]$

\bigskip

$\bullet$ Consistent with HOMFLY -- Jones ($N=2$) -- Alexander ($N=0$) reductions
(by definition)

\bigskip

$\bullet$ Consistent with Heegard-Floer polynomials $HF_R(q|t)$ \cite{EGhf}

\bigskip

$\bullet$ Consistent with superpolynomials, evaluated by the
sums of paths on $2d$ lattices ($q,t$-Catalan numbers) \cite{EGcat}

\bigskip

$\bullet$ Reproduces $P_{[2]}^{[2,3]}$ of \cite{ASh},
but does not reproduce Hopf link superpols $P_{[2],[1^s]}^{[2,2]}$
of \cite{GIKV} and \cite{AK} (because of the different choice of unknot superpolynomial)

\subsection{Generalizations \cite{DMMSS}}

\noindent

$\bullet$ Higher non-fundamental representations $R\neq [1^{|R|}]$.\\
The basic issue here is the choice of unknots:\\
if one takes MacDonald dimension $M_R^*$ as the answer, then
already its simplest building block $\frac{Aq-(Aq)^{-1}}{tq-(tq)^{-1}}$
is not a polynomial, even if $A=t^N$.

\bigskip

$\bullet$ {Link invariants}:\\
Do superpolynomials exist at all for toric links?\\
There are still different opinions on what can be sacrificed:

the best is to release the polynomiality condition,

alternatively there are attempts to release the
positivity condition \cite{AK}.\\
Split $W$-evolution for torus links,

breaks {\it both} these conditions,

does not reproduce the answers from \cite{GIKV},

still it seems to be the right prescription.\\
Even if evolution originates from the modified unknots,

which are {\it forced} to be  {\it polynomial},

the positivity condition is broken.

\bigskip

$\bullet$ {Non-toric knots:}\\
The main approach is outlined in sec.3.8 of \cite{DMMSS} and \cite{knMMM12},
again one can use series of knots.\\
Potentially successful example of this approach is provided by the lifting
$5_2 \ \longrightarrow\ 10_{139}$.\\
Breakdown of positivity for evolution of $4_1$ implies that the superpolynomials
for the {\it composite} knots \cite{knMMM12} \\
do {\it not} possess the positivity property.

\section{Conclusion}

To conclude, $\beta$-deformation brings us into a completely new world,
where familiar {\it structures} are non-trivially deformed,
but final {\it answers} have absolutely straightforward generalizations:
one can easily write them down and validate, without knowing the reasons
for them to be true.
This combination makes the subject so interesting and important
for development of string theory methods.
This development can prove to be rather technical at the algebraic
(group-theory) side of the story, but it definitely looks conceptual
at the geometric side.

\section*{Acknowledgements}

I am indebted for discussions and help to my colleagues, collaborators and friends
A.Alexandrov, V.Dolotin,
P.Dunin-Barkovsky, D.Galakhov, E.Gorsky, S.Gukov, S.Kharchev, A.Marshakov, A.Mironov,
And.Morozov, S.Na\-tanzon, A.Po\-politov, Sh.Shakirov,
A.Slep\-tsov, A.Smirnov, A.Zabrodin,  Y.Zenkevich
and especially to Maxim Zabzine.
My work is partly supported by
the Ministry of Education and Science of the Russian Federation under contract
14.740.11.081,  by RFBR grant 10-02-00499 and by the joint grants 11-02-90453-Ukr,
12-02-91000-ANF, 12-02-92108-Yaf-a, 11-01-92612-Royal Society.

\end{document}